\documentclass[twoside,10pt]{article}
\usepackage{amsmath,amssymb}
\usepackage{xspace}
\usepackage{tam13}

\title{\centerline{Heavy Quarks, Origin of Mass, %} \vskip0.2cm
and CP Violation for Universe}
}
\shorttitle{Mass \& CPV from Heavy Quarks}
\authors{George W.S.~Hou}
\shortauthors{G.W.S.~Hou}
\affiliations{
Department of Physics, National Taiwan University, Taipei, Taiwan 10617
}

\abstract{
A scale-invariant "Gap Equation" is constructed for 
chiral quark $Q$ by Goldstone, or $V_L$, exchange, 
where massless input is guaranteed by gauge invariance. 
A numerical solution is found for Yukawa coupling $\sim 4\pi$. 
In turn, because this gap equation is scale invariant, 
the strong coupling solution is compatible with a 126 GeV dilaton, 
which would be a true messenger from higher energies. 
Some possible phenomena pertaining to heavy chiral quarks
at few TeV scale is offered.
Adding this heavy quark sector may provide enough CP violation
for generating the matter dominance of the Universe. 

}

\begin{document}
\maketitle

\section{\underline{Introduction}}

The Holy Grail of particle physics appears to have been found
in 2012: with the triumphal at CERN on July 4th, Fabriola Gianotti
(then ATLAS spokesperson) became a TIME magazine ``Person of the Year'',
and the Higgs boson was proclaimed ``Breakthrough of the Year'' by \emph{Science}.

Let's go back to the original.
While citing Anderson's insight on the plasmon phenomena,
the landmark 1964 paper~\cite{Higgs64} by Higgs made clear that,
for spontaneously broken gauge theories,
the gauge bosons corresponding to the broken symmetry
turn massive; in the limit of vanishing gauge coupling,
the longitudinal modes of these gauge bosons revert to
the massless Goldstone modes of the broken symmetry.
However, citing BCS theory for superconductivity,
the end note is that the symmetry-breaking scalar field
need not be elementary, but could be fermion bilinears.

Half of the 2008 Nobel Prize went to Nambu,
 ``for the discovery of the mechanism of
   spontaneous broken symmetry (SSB) in subatomic physics''.
As the prize was received together with Kobayashi and Maskawa
for the CP violation mechanism of the Standard Model (SM),
I was asked to review~\cite{Hou09} the 2008 prize at the FPCP 2009 conference.
Even before the prize announcement, however, through my own experimental
and theoretical interests, I became fascinated by the prospect or
possibility that,
\begin{center}
\emph{Could electroweak symmetry breaking (EWSB) be due to \\
$b'$ and $t'$ quarks above unitarity bound $\sim 500$--$600$ GeV?}
\end{center}
Here, $b'$, $t'$ are 4th generation quarks.
The thought, based on the Nambu--Jona-Lasinio (NJL) model~\cite{NJL},
goes back to Nambu in the late 1980s.
So let me quote Nambu's Nobel Lecture~\cite{Nambu08} comments on the Higgs mechanism:
\begin{itemize}
 \item ``I thought the plasma and the Meissner effect had established it.''\hskip1cm
       [he was ahead of even Anderson]
 \item ``I should have paid more attention to the Ginzburg--Landau theory
       which was a forerunner of the present Higgs description.''
 \item Citing $^3$He superfluidity, nucleon pairing in nuclei as BCS-like SSB,
       he questions fermion mass generation in EW sector,
        ``my biased opinion, there being other interpretations to
          the nature of the Higgs field.''
\end{itemize}

The last point, often glossed over in textbooks, touches on 
not only possible ``bosonization by fermion pairing'' for the Higgs field,
but the issue of fermion mass generation in the Standard Model (SM),
\begin{equation}
 m_f = \frac{1}{\sqrt2}\, \lambda_f\, v,
\end{equation}
where $\lambda_f$ is the Yukawa coupling and $v$ the vacuum expectation value.
This is quite similar to $m_V = \frac{1}{2}\, g\,v$,
which is the original Higgs mechanism.
The gauge coupling $g$, however, is a dynamical concept arising from symmetry principles.
Is Eq.~(1) fortuitous, or intended? 
If the latter, then why do we have an unwieldy scatter of 
9 quark and charged lepton masses over 6 orders of magnitude?
As the ``God Particle'', the Origin of Fermion Masses through
a plethora of ``random'' Yukawa couplings is not quite seemly.

Starting from Eq.~(1) but with strong Yukawa coupling,
we offer some thoughts and observations on \emph{dynamical} EWSB.

\section{\underline{NJL, Top, Scaled-up QCD}}

\subsection{Top-quark Condensation}
\label{ss:BHL}

Let us retrace some history of heavy quark induced dynamical symmetry breaking (DSB).
As bounds on the top quark mass rose towards the end of the 1980s,
following a suggestion by Nambu,
Bardeen (son of the bi-laureate John Bardeen of BCS), Hill and Lindner
proposed~\cite{BHL89} DSB through top condensation, 
i.e. $\langle \bar tt\rangle \neq 0$,
by analogy with BCS theory. Borrowing the language, they formulated a
``Gap Equation" through the NJL model~\cite{NJL}.
Unfortunately, the top quark turned out not massive enough.

\begin{figure}[htb]
\centering
\includegraphics[width=90mm]{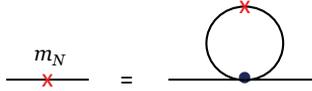}
\vskip-5cm
\caption{Gap equation of NJL model for nucleon mass generation.
\label{f:gapNJL}}
\end{figure}

\subsection{NJL Model and Gap Equation}
\label{ss:NJL}

The NJL model~\cite{NJL} was an explicit realization of earlier
insights by Nambu. By analogy to the appearance
of an ``energy gap" in BCS superconductivity, NJL suggest that
the nucleon mass ``arises largely as the self-energy of a
primary fermion field''. Considering an effective 4-fermi interaction,
by joining the two legs to form a self-energy bubble, they ask whether 
the fermion mass can be generated self-consistently this way
(Fig.~\ref{f:gapNJL}), then showed explicitly that the answer is the positive.
Since fermion mass generation violates chiral symmetry,
``there arise automatically pseudoscalar zero-mass bound states
of nucleon-antinucleon pair'' which is the pion
 (now a bound state of $q \bar q$ pair).
This is $m_N$ generation as SSB through the dynamical 4-fermi interaction.

The NJL model picturised in Fig.~\ref{f:gapNJL} forms a Gap Equation, 
with the (red) cross \textsf{X} representing the self-energy.
Note that, identifying it with the bubble,
this loop function does not depend on the external momentum $p_N$
of the nucleon. Thus, \textsf{X} is nothing but a constant, i.e. $m_N$,
and the same holds inside the loop on the r.h.s. of the equation.
On one hand, one could by iteration generate an infinite number of diagrams
(the pion can be generated analogously by considering $N\bar N \to N\bar N$ scattering),
on the other hand, one simply has the gap equation
\begin{eqnarray}
 m_N &=& \frac{N_C}{8\pi^2}G\int_0^{\Lambda^2} dq^2\, q^2 \frac{m_N}{q^2 + m_N^2} \nonumber \\
     &=& \frac{N_C}{8\pi^2}G\Lambda^2
         \left(1 - \frac{m_N^2}{\Lambda^2} \log\left(1 + \frac{\Lambda^2}{m_N^2}\right)
         \right)m_N,
 \label{e:mN}
\end{eqnarray}
where $G$ here is the 4-fermi coupling and $\Lambda$ is the cutoff.
Factoring out $m_N$ (because of $p_N$-independence), one has
\begin{eqnarray}
 1 - \frac{G_{\rm crit}}{G} &=& \frac{m_N^2}{\Lambda^2} \log\left(1 + \frac{\Lambda^2}{m_N^2}\right),
 \label{e:NJL}
\end{eqnarray}
which admits a solution for $G > G_{\rm crit}$, with
\begin{eqnarray}
 {G_{\rm crit}} &=& \frac{8\pi^2}{N_C \Lambda^2}.
 \label{e:Gcrit}
\end{eqnarray}
One eventually trades the parameters $G$ and $\Lambda$
for the physical $f_\pi$ and $m_N$.

\subsection{Scaled-up QCD}
\label{ss:TC}

It is amazing to think that the NJL model predates quarks, QCD, and all that ...
It can be formulated in terms of quarks rather than nucleons
(the factor $N_C$ in Eq.~(\ref{e:mN})).
Within QCD, it is still not fully understood what ``is'' the
pointlike effective 4-fermi interaction.
But a mainstream approach to dynamical EWSB is, of course, scaling up QCD.

Compared with QCD (or Quantum Chromodynamics), the actual
nonAbelian gauge theory of the strong interactions that generates $m_N$ by DSB,
TechniColor (TC) is but a mock-up in scale by a factor of 2000.
We know that ``technicolor'' is not as good as true color.
There are multiple associated problems ... presumably resolved by ``Walking''
(WTC), or tuning the effective $N_{\rm TF}$.
But, take for example the LatKMI collaboration formed just two years ago
at the Kobayashi--Maskawa Institute~\cite{KMI}, serious computing money is involved,
while the Yale group is certainly not smaller.

Our approach would not rely on analogy with QCD.
While invoking pairing, it would not be just NJL.

\section{\underline{\boldmath From $Q\bar Q$ Scattering to Heuristic ``Gap Equation''}}

Physics is an empirical Science. A chain of empirical observations
during 2009--2011 lead me to a Gap Equation that was not quite NJL-like.

\subsection{Empirical 1:  Reverse-engineer ``Goldstone'' and ``Yukawa'' }
\label{ss:Emp1}

Since 2006-7, I started to think about what physics
could a small, far away group do in CMS at the LHC.
The answer was to set up the 4th generation search program.
Since 2008, as bounds rose, I relearned the aforementioned
Yukawa-induced condensation.
But around 2009, I saw that~\cite{Hou:2012az}, because we live and work in
a broken massive world, the purely left-handed gauge coupling,
well tested by LEP experiments, actually contains the usual Yukawa coupling.

\begin{figure}[t!]
\begin{center}
\vskip-2.1cm
 \includegraphics[width=90mm]{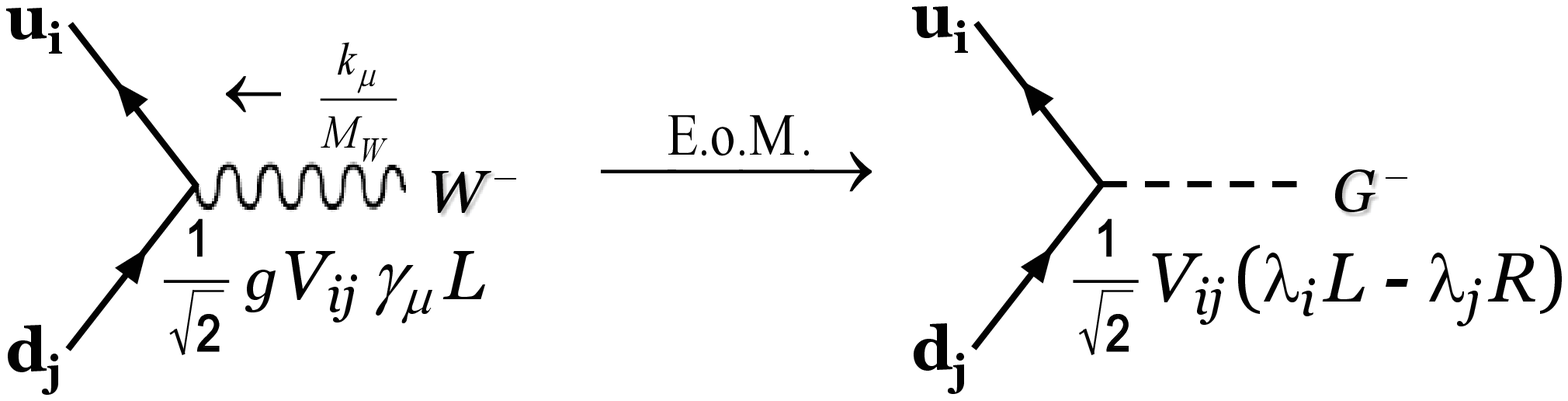}
\vskip-2.3cm
\caption{Yukawa coupling
 from purely left-handed gauge coupling.}
 \label{f:YukGau}
\end{center}
\end{figure}

As illustrated in Fig.~\ref{f:YukGau},
contracting the gauge vertex with $k_\mu/M_W$,
which propagates the longitudinal mode of weak boson $W$
($g'$ coupling already turned off), by applying the E.o.M.
(the equation of motion, i.e. nothing but Dirac equation),
we arrive at the r.h.s.,
where the gauge coupling $g$ cancels out the $g$ in $M_W = gv/2$,
and we arrive at the standard definition of Yukawa coupling
\begin{equation}
\lambda_Q \equiv \frac{\sqrt{2} m_Q}{v},
 \label{e:Yuk}
\end{equation}
which is the coupling of $W_L$, or Goldstone boson (one can take $g \to 0$ limit!),
to quarks, and \emph{one and the same as Eq.~(1)}.
Given that the left-handed gauge coupling is now empirically established
by LEP to the per mille level,
and given that we only applied the Dirac equation in an empirically all-massive world,
the Yukawa coupling that mixes left- and right-handed fermion components
is empirically established!
Given that the Higgs boson is not mentioned in the above observation,
we assert
\begin{center}
\emph{ Yukawa Coupling $\lambda_Q$ of Goldstone mode $G$ is experimentally\\
 established, independent of existence of the Higgs boson $H$.}
\end{center}
In SM, $G$ and $H$ form a complex scalar doublet field.

\subsection{Empirical 2:  ``Unitarity Bound'' Violation via Long-distance?
}
\label{ss:Emp2}

Reinforced by Nambu's prize announcement in October 2008,
I was conditioned to think that $Q\bar Q$ condensation is
via an NJL-like mechanism, i.e. via 4-quark or contact interaction
between heavy quarks.

\begin{figure}[t!]
\begin{center}
\vskip-1.8cm
 \includegraphics[width=82mm]{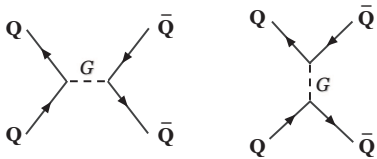}
% \hspace{5mm}
% \includegraphics[width=30mm]{ggtoV8g.eps}
\vskip-2.1cm
\caption{$Q\bar Q$ scattering via $t$- and $s$-channel
 Goldstone $G$ exchange.}
 \label{scatt}
\end{center}
\end{figure}

As the bounds on 4G quarks rose, I started to pay attention to
unitarity bound violation (UBV), the notion that
for $m_Q \gtrsim 550$ GeV, $Q\bar Q$ (and $QQ$) scattering diverge at high energy.
This, of course, is nothing but the onset of Strong Coupling in Eq.~(5),
and need not be feared.
However, starting in 2009, I became puzzled.
Tracing the root of UBV, the dominant $Q\bar Q$ scattering is
actually through $t$-channel Goldstone $G$, or longitudinal $W_L$,
exchange,\footnote{
The leading UBV is the repulsive $s$-channel $G$
(r.h.s. of Fig.~\ref{scatt}) exchange.
But one may question whether such a Goldstone $G$ can be sustained at high $\sqrt{s}$.
}
as seen in Fig.~\ref{scatt}.
This is because, for large $m_Q$, its Yukawa coupling from Eq.~(5)
is the largest coupling, larger than weak coupling $g$,
larger even than strong coupling $g_s$.
The Goldstone particle can be viewed as having mass $M_W$.
We then have the puzzle that the strongest UBV term is carried
by relative long-distance (LD) interaction, with scale $1/M_W$,
compared to the localization scale $1/m_Q$ for the heavy quark.
\begin{center}
\emph{The leading UBV for heavy $Q\bar Q$ scattering does not appear NJL-like.}
\end{center}

\subsection{Empirical 3:  ``Unitarity Bound'', $Q\bar Q$ Scattering
 \ $\Rightarrow$ \ Self-Energy
}
\label{ss:Emp3}

As I puzzled over the UBV problem with ``$Q\bar Q$ scattering at long distance",
in 2010 I came to an ``empirical'' Gap Equation that differs from NJL.

In summer 2010, when passing through Munich on my way to ICHEP held in Paris,
I had the occasion to discuss with Felipe Llanes-Estrada about my troubles.
Drawing the left-hand part of Fig.~\ref{scatt} on the blackboard,
all of a sudden, I jumped. Connecting the upper two lines,
i.e. from $Q$ to $\bar Q$, I exclaimed that this is a self-energy.
Could this not form a Gap Equation and self-generate? The mindset changed.

\begin{figure}[t!]
\begin{center}
\vskip0.4cm
 \includegraphics[width=80mm]{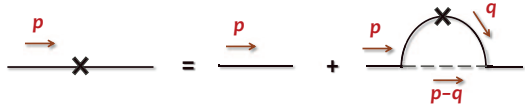}
\vskip-4.5cm
\caption{Gap equation by Goldstone exchange for large Yukawa coupling
         in the ladder approximation, with $m_0 = 0$.}
 \label{GapYuk}
\end{center}
\end{figure}

The ``Gap Equation'' is shown in Fig.~\ref{GapYuk}.
Ignoring for now the first term on the r.h.s. (explained later),
Fig.~\ref{GapYuk} reverts to the Gap Equation of NJL model,
Fig.~\ref{f:gapNJL}, \emph{if} one could shrink the Goldstone propagator to a point.
But the existence of this propagator illustrates the difference:
the external momentum $p$ of the ``heavy quark'' enters the loop,
as one can see from the momentum labels.
This means that the cross \textsf{X}, which represents
the self-energy function on the l.h.s., depends on $p$,
hence when reinserted into the loop on the r.h.s.,
is considerably more complicated than the NJL model.

The internal momentum $q$, summed over for the self-energy,
corresponds to all possible momenta exchange carried by
the Goldstone in the scattering diagram of Fig.~\ref{scatt}.
Thus, the self-energy corresponds to integrating over
all possible momenta, and UBV implies that this integration
cannot go on for all scales.

With rapid accumulation of data, 
apprehension rose at the LHC in 2011: ``1--2--3 (TeV), No New Physics''.
As was already apparent by summer 2011,
where the above slogan is paraphrased from,
the naive SUSY scale, new gauge boson mass, and excited quark mass, 
respectively, cannot be less than 1, 2, 3 TeV;
bounds have only become more stringent since.
This made a strong impression on me.
With heavy chiral quark (4G) mass bounds reaching and 
breaching the unitarity bound,
the absence of any New Physics implied that
one could sum over $q$ in Fig.~\ref{GapYuk} up to a rather high scale.
The question now is then:
\begin{center}
\emph{Can Goldstone exchange with large Yukawa coupling \\
$\lambda_Q$ generate $m_Q$ above UBV self-consistently?}
\end{center}
Publishing the speculations~\cite{Hou:2012az} that lead to
a precursor Gap Equation to Fig.~\ref{GapYuk}, 
I embarked on this investigation with collaborators.

%\section{Scale-Invariant (Strongly Interacting) QED:  Setup}

\section{\underline{Numerical Solution of Strong Yukawa Gap Equation}}

\subsection{Momentum-dependent Self-energy and Strong QED}
\label{ss:SIQED}

We have already emphasized the momentum-dependent self-energy
for the Gap Equation of Fig.~\ref{GapYuk}, which differs from
the NJL gap equation of Fig.~\ref{f:gapNJL}.
In fact, Fig.~\ref{GapYuk} bears much similarity to
strongly interacting, scale-invariant QED (SIQED),
if one replaces the Goldstone $G$ by the photon $\gamma$.
The curiosity for SIQED is, if one started with $m_0 = 0$
for the electron (hence scale invariant),
could one generate a finite electron mass
(hence dynamically break chiral symmetry)
by strong coupling $\alpha = e^2/4\pi$?
The answer is that this is possible,
with a critical coupling of $\alpha_{\rm crit} = \pi/3$.

Although we will follow the methodology~\cite{FK76} developed for solving SIQED,
to save space, we will not go through the steps,
except to note that our strong Yukawa Gap Equation is more ``natural":
$m_0 = 0$ is enforced by gauge invariance in a chiral gauge theory setting.
On the other hand, the Yukawa Gap Equation has one
peculiarity over SIQED: starting from the empirical Goldstone boson $G$,
we incorporate it in a self-energy loop, such that its own existence
can be justified if it generates the heavy quark mass $m_Q$,
thereby breaking the (gauge) symmetry dynamically.
This is a form of ``bootstrap'' that Nambu advocated~\cite{Nambu88}.

\subsection{Mathematical Formulation}
\label{ss:Formulation}

The chief thing learned from SIQED is that,
because of the momentum dependence of the self-energy,
one should write
\begin{equation}
S(p)^{-1} = A(p^2)\, /\!\!\!\!p - B(p^2),
 \label{SQ-1}
\end{equation}
keeping the wave-function renormalization term $A(p^2)$.
Mass is now defined as ``$B(p^2)/A(p^2)$'' on mass-shell.
The self-energy is then $S_0(p)^{-1} - S(p)^{-1}$
where $S_0(p)^{-1} = /\!\!\!\!p - m_0 = /\!\!\!\!p$; 
since $m_0 = 0$, one recovers Fig.~\ref{GapYuk}.
Inserting $S(q)$ into the \textsf{X}
on r.h.s. of Fig.~\ref{GapYuk},
assuming the Goldstone propagator is a simple $1/q^2$ (Goldstone condition)
while $\lambda_Q$ is just a number, one gets
the coupled integral equations,
\begin{eqnarray}
B(p^2) &=& \frac{3\lambda_Q^2}{2}
\int \frac{d^4q}{i(2\pi)^4}
\frac{1}{(p-q)^2}
\frac{
B(q^2)
 }{q^2 A^2(q^2) - B^2(q^2)}\nonumber \\
&-& \frac{\lambda_Q^2}2
 \int \frac{d^4 q}{i(2\pi)^4}
\frac{1}{(p-q)^2-m_h^2}
\frac{
B(q^2)
 }{q^2 A^2(q^2) - B^2(q^2)},
 \nonumber \\
% \label{Bp2}
%\end{eqnarray}
%
%and
%
%\begin{eqnarray}
A(p^2) &=& 1 + \frac{3\lambda_Q^2}{2p^2}
\int \frac{d^4q}{i(2\pi)^4}
\frac{p\cdot q}{(p-q)^2}
\frac{
A(q^2)
 }{q^2 A^2(q^2) - B^2(q^2)}\nonumber \\
&+& \frac{\lambda_Q^2}{2p^2}
 \int \frac{d^4 q}{i(2\pi)^4}
\frac{p\cdot q}{(p-q)^2-m_h^2}
\frac{
 A(q^2)
 }{q^2 A^2(q^2) - B^2(q^2)},
 \label{p2Ap2}
\end{eqnarray}
where we have included the Higgs scalar contribution.
We are interested in the absence of the $h$ term
(which we would return to discuss).
But if one considers the $h$ contribution,
given that $m_h$ is light, we can set it to zero,
and the Higgs term just modulates the coefficients of the integrals.

After angular integration and Wick rotation, one gets
\begin{eqnarray}
B(x) = \kappa_b
\left(
\frac1{x} \int_{\Lambda_{\rm IR}^2}^x dy \frac{y B(y)}{y A^2(y)+ B^2(y)} %\right.
%\nonumber \\  && \left. \quad\;\ \
 + \int_x^{\Lambda^2} dy \frac{B(y)}{y A^2(y)+B^2(y)} \right),
 \label{Bx} \\
A(x) = 1 + \kappa_a
\left(
\frac1{x^2} \int_{\Lambda_{\rm IR}^2}^x dy \frac{y^2 A(y)}{y A^2(y)+ B^2(y)} %\right. %\nonumber \\ && \left. \quad\quad\;
+ \int_x^{\Lambda^2} dy \frac{A(y)}{y A^2(y)+B^2(y)}
\right),
 \label{Ax}
\end{eqnarray}
where $\Lambda_{\rm IR}$ and $\Lambda$ are the IR and UV cutoffs, respectively,
and
\begin{eqnarray}
\kappa_b = 2\kappa_a
 = \frac{3\alpha_Q}{8\pi} \ \ {\rm (no\ Higgs)}; \quad
\kappa_b = \kappa_a %= \frac{\lambda_Q^2}{16\pi^2}
 = \frac{\alpha_Q}{4\pi} \ \ {\rm (massless\ Higgs)}.
 \label{kappa}
\end{eqnarray}

For the case of SIQED, choosing the Landau gauge
gives $A(p^2) = 1$, and the coupled equations
reduce to a single integral equation, which can be
transformed to a differential equation with boundary conditions.
For our case, one cannot avoid the coupled equations,
but one can still differentiate Eqs.~(\ref{Bx}) and (\ref{Ax}),
and after some manipulations, arrive at
\begin{eqnarray}
x B^{\prime\prime} + 2 B^\prime + \frac{\kappa_bB}{x A^2 + B^2} = 0 , \quad
% \label{xBpp} \\
x A^{\prime\prime} + 3 A^\prime + \frac{2\kappa_aA}{x A^2+B^2} =0,
% \label{xApp}
 \label{xBApp}
\end{eqnarray}
with boundary conditions
\begin{eqnarray}
&&\left.B^\prime(x)\right|_{x=\Lambda^2_{\rm IR}} = 0,
\quad  \left[xB^\prime(x)+B(x)\right]|_{x=\Lambda^2} = 0, \\
&&\left.A^\prime(x)\right|_{x=\Lambda^2_{\rm IR}} = 0,
\quad  \left[\frac{x}{2} A^\prime(x)+ A(x)\right]|_{x=\Lambda^2} = 1.
\end{eqnarray}
Redefining $p^2 = x = e^{2t}$,
one gets
\begin{equation}
\ddot{B} + 2 \dot{B} + \frac{4\kappa_bB}{A^2+B^2 e^{-2t}}=0, \quad
\ddot{A} + 4 \dot{A} + \frac{8\kappa_aA}{A^2+B^2 e^{-2t}}=0,
\end{equation}
\begin{eqnarray}
&&\dot{B}(t_{\rm IR}) = 0, \;\ \quad \dot{B}(t_{\rm UV}) + B(t_{\rm UV}) =0, \\
&& \dot{A} (t_{\rm IR}) = 0,\quad \frac14\dot{A}(t_{\rm UV}) + A(t_{\rm UV}) = 1,
\end{eqnarray}
where dot represents $t$-derivative,
and $e^{t_{\rm UV}} = \Lambda_{\rm UV} = \Lambda$,
$e^{t_{\rm IR}} = \Lambda_{\rm IR}$.

Due to scale invariance,
the differential equations are invariant under
\begin{eqnarray}
x \to a^2 x   \ \ (t \to t+ \log a); \quad
\Lambda_{\rm UV,IR} \to a \Lambda_{\rm UV,IR}; \quad
B \to a B, \ A\to A.
\label{scale_inv}
\end{eqnarray}
Thus, the solutions depend only on
$\Lambda_{\rm UV}/\Lambda_{\rm IR}$ ($=e^{t_{\rm UV}- t_{\rm IR}}$) and
$m_{\rm dyn} \equiv B(t_{\rm IR})/A(t_{\rm IR})$
for given $\kappa_a$ and $\kappa_b$,
and $m_{\rm dyn}$ is a kind of integration constant.
We find that the most important feature of the solutions is that
only special discontinuous values of $\kappa_a$ and $\kappa_b$
are allowed for given B.C.

\subsection{Numerical Results}
\label{ss:Result}

The solution is found numerically, and
\begin{eqnarray}
\kappa_b^c \simeq 1.4, \ \ (\kappa_b = 2\kappa_a = 3\alpha_Q/8\pi); \quad
\kappa_b^c \simeq 13.7,\ \ (\kappa_b = \kappa_a = \alpha_Q/4\pi),
\end{eqnarray}
corresponding to
\begin{eqnarray}
&&\lambda_Q^c \simeq 12,  \qquad\; (\kappa_b = 2\kappa_a = 3\alpha_Q/8\pi,
 \ {\rm i.e.\ no\ Higgs})
 \label{lamcr_infty} \\
&&\lambda_Q^c \simeq 46,  \qquad\; (\kappa_b = \kappa_a = \alpha_Q/4\pi,
 \ {\rm i.e.\ massless\ Higgs})
 \label{lamcrHX_infty}
\end{eqnarray}
much higher for the latter case.
These values are extracted in the
large $\Lambda_{\rm UV}/\Lambda_{\rm IR}$
and $\Lambda_{\rm UV}/m_{\rm dyn}$ limit,
and $c$ stands for ``critical".
Note that for the artificial case of $\kappa_a = 0$ (i.e. $A = 1$),
the critical value is $\lambda_Q^c \simeq 5.1$,
hence the effect of $A$ or wave function renormalization
is nontrivial.

Eq.~(\ref{lamcr_infty}) translates, via the empirical Eq.~(5) 
with $v = 246$ GeV, into
\begin{equation}
 m_Q^c > 2.1\ {\rm TeV}. \quad\quad {\rm (No\ Higgs)}
 \label{mQcr_infty}
\end{equation}
In contrast, Eq.~(\ref{lamcrHX_infty})
would lead to a much larger value of $m_Q^c =  8.1$ TeV.

If the result of Eq.~(\ref{mQcr_infty}) is already astonishing,
we note that for the Gap Equation to be self-consistent,
then $\Lambda < 2m_Q$ should be kept.
But in our numerical result, we have integrated to $\Lambda \to \infty$.
If we integrate to $\Lambda < 2m_Q$ self-consistently,
then $\lambda_Q^c$ is raised from 12 to 17.7,
and $m_Q \sim 3$ TeV, which is depressingly large.
Putting back the light Higgs would only make matters worse, 
as we have already noted.

Can $\lambda_Q^c$ be made lower? Possibly, because
strong Yukawa implies tightly bound heavy ``mesons''~\cite{EHY11},
which can add to the Gap Equation with NJL-like,
but momentum-dependent self-energy terms.
Since we have not solved the bound state problem,
this remains a conjecture.
We note that $\lambda_Q^c \sim 12$--17.7 is
reminiscent of the pion-nucleon system, 
where $\lambda_N \simeq 14$,
a point we will come back to later.

\section{\underline{Viability?   VBF Arbitration}}

\subsection{Light ``Higgs'', or is it the Dilaton?}
\label{ss:light-h}

Let us finally face the question: What about the 126 GeV boson?
The 4th generation is perceived as meeting its fate with the July 4th observation.
The short answer to the question, all things considered, 
is that ``It (the 126 GeV boson) would have to be dilaton!''

Even when the hint for the 126 GeV boson first appeared in 2011,
theorists have cautioned that what may appear as a light Higgs boson
could be an ``imposter'', such as a dilaton.
A variant version would be the warped extra dimension ``radion''.
Let me quote the following statement~\cite{EP12}, drawn from the WTC camp:
``holographic techni-dilaton  gives as good a fit as SM Higgs''.
But honestly, my first reaction was that, 
\emph{it is really hard to believe we are seeing a dilaton ``just now''},
and I have been depressed since July 4th, 2012.
Just to remind you, the dilaton ${\cal D}$ is the Goldstone boson of 
SSB of Scale Invariance, so it felt rather ``theoretical'' to me.

The authors of Ref.~\cite{EP12}, where the above quote is drawn,
add the parameters $a \equiv v/F$, $c_g$ and $c_\gamma$ to their simple fit.
The dilaton scale $F$ suppresses all couplings to
$VV$ and $\bar ff$ final states,
while $c_g$ and $c_\gamma$ are free parameters of the 
${\cal D}gg$ and  ${\cal D}\gamma\gamma$ couplings over the SM value,
which are determined by the $\beta$ function of QCD and QED, respectively,
because of the trace anomaly of the energy-momentum tensor.
We can now use these to see how the above quote is drawn,
which we apply for 4G.

The reason that 4G is viewed as ruled out by the
126 GeV ``Higgs'' is because, having the $b'$ and $t'$ in
the loop in addition to the top, $gg \to h$
production cross section would be enhanced by a factor of 9. 
This enhancement is inconsistent with the observed $ZZ^*$ production
within SM.
If the observed 126 GeV ``Higgs'' is in fact a dilaton,
then the factor of 9 is replaced by $|c_g|^2$,
but the ${\cal D}ZZ$ coupling gets suppressed by $a = v/F$,
and the outcome could be what we see, i.e. still consistent with SM.
Likewise, $gg \to {\cal D} \to \gamma\gamma$ would be
modulated by $|c_gc_\gamma|^2$,
implying a suppression of $c_\gamma$, which occurs for 4G. 
Given the present experimental precision,
this general dilaton view is not ruled out,
as the two channels of observation are precisely
$ZZ^*$ and $\gamma\gamma$ via gluon-gluon fusion.

To rule out the dilaton, 
one must measure the $VV$ coupling of the observed boson, 
i.e. to establish VBF production is consistent with SM,
versus the $v^2/F^2$ suppression in the dilaton case.
Current sensitivity for the CMS and ATLAS experiments~\cite{LP2013} is 
at the $2\sigma$ level, with ATLAS claiming 3$\sigma$ observed significance.
It seems that VBF production of the 126 GeV boson 
cannot be established unequivocally before the 13 TeV run.

\subsection{Dilaton Self-consistency for Yukawa Gap Equation}
\label{ss:dilaton-consistent}

We now comment on how the dilaton fits into our Gap Equation framework.
Recall that our Yukawa Gap Equation is scale invariant.
In fact,we have used scale invariance in Eq.~(17) to
help find the numerical solution.
One may think that dynamical mass generation as demonstrated by
the numerical solution would also break scale invariance.
But one has to address the true source of scale invariance violation,
which we think is the ``\emph{Theory of Yukawa Coupling}'',
i.e. the theory for $\lambda_Q$, which is still lacking.
We have treated it only as a number in our Gap Equation.
We note that our empirical-based Gap Equation
cannot be integrated beyond $2m_Q$, hence one is shielded from the UV theory.

Actually, assuming the observed 126 GeV boson is the dilaton 
is self-consistent within our Gap Equation approach. 
We have found that, if we assume the 126 GeV boson is the
Higgs boson $h$ in SM, then the needed $\lambda_Q$ to achieve DSB
is untenably large. But if we use ${\cal D}$ instead,
then the second terms for $B$ and $A$ in Eq.~(7) are
suppressed by $v^2/F^2$, which can be ignored 
at the same level as we have ignored $g^2$ and $g_s^2$ effects.

\section{\underline{Curious Phenomenology; and Fermi-Yang Redux}}

We now draw on analogy to the $\pi$-$N$ system to
discuss similarities, and dissimilarities, with the $G$-$Q$ system.

\subsection{\boldmath $Q\bar Q \to nV_L\,$?}
\label{ss:QQtonVL}

Let us start with a quiz: How does $p\bar p$ annihilate?
An easy response is ``into photons'', by simply analogy with $e^+e^-$.
But the proton is strongly interacting. 
So, a likely second response is ``into gluons''.
However, with available energy of no more than a GeV,
one cannot clearly identify a gluon.

\begin{figure}[htb]
\centering
%\vspace{7mm}
 {\includegraphics[width=80mm]{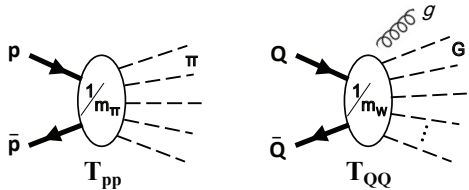}
 }
\vskip-35mm
\caption{
Annihilation of $p\bar p \to n\pi$, and the analog of $Q\bar Q \to nG$,
where $G = V_L$ is the Goldstone boson.
The gluon line indicates shedding of color.
} \label{QQ-to-nG}
\end{figure}

Facing this question in early 2012, I reasoned that 
$p\bar p$ annihilate into mesons. 
Since all mesons end up as pions, the simple conclusion is that: 
$p\bar p$ annihilate into a number of pions.
But by Spring 2012, when I read up on the literature, 
it was totally fascinating: it is \emph{experimentally known}
that $p\bar p$ annihilate via $n\pi$ ``fireball'' (see Fig.~\ref{QQ-to-nG}),
\begin{itemize}
\item Size of order $1/m_\pi$;, with temperature $T \simeq 120$ MeV;
\item Average number of emitted pions $\langle n_\pi \rangle \simeq 5$;
\item A soft-pion $p_\pi^2/E_\pi^2$ factor
 modulates the Maxwell--Boltzman distribution for the pions.
\end{itemize}

Recalling some strange statement from nuclear physics: 
``the strong $g_{\pi NN}$ coupling extracted from 
Born approximation is above 10, even though this makes 
the approximation dubious'',
it struck me further that
\begin{eqnarray}
g_{\pi NN} \simeq \lambda_{\pi NN} \equiv \sqrt2 \frac{m_N}{f_\pi} \simeq 14,
\end{eqnarray}
within experimental accuracy.
Here, $\lambda_{\pi NN}$ is defined as the ``Yukawa coupling''
of the Goldstone pion to the nucleon,
and is found to be of similar strength to $g_{\pi NN}$ coupling,
both of order 14, and above the ``Naive Dimensional Analysis''
strong coupling of $4\pi$.
This realization made me more comfortable with
Eq.~(21), and our finding of $\lambda_Q^c \simeq 12$--$17$
from our numerical solution to the Gap Equation.

From Eqs.~(5), (21) and (22), I could draw~\cite{Hou:2012df} 
an intriguing analogy: the heavy $Q\bar Q$ would 
annihilate to $nG$, where $G$ is the $V_L$ or Goldstone boson of EWSB! 
This is shown on the r.h.s. of Fig.~\ref{QQ-to-nG},
where in general one would need to radiate a (relatively soft) gluon
to shed color.
By analogy, the thermalization region is of size $1/m_W$
with temperature $T$, which should be at weak scale.
Taking $T = 2v/3 \simeq 160$ GeV for example,
we find
\begin{equation}
 \langle |p_G| \rangle \sim 310\ {\rm GeV},
 \label{pG}
\end{equation}
and for $m_Q = 1\ (2)$ TeV, or $2m_Q = 2\ (4)$ TeV,
this corresponds to
\begin{equation}
 \langle n_G \rangle \sim 6\ (12).
 \label{nG}
\end{equation}
with a multiplicity distribution in analogy with $p\bar p$ annihilation.
Aided by Yukawa boundstates,
this phenomenology may yet be revealed at the 13--14 TeV runs of the LHC.

\subsection{Fermi--Yang Model Redux}
\label{ss:FermiYang}

The analogy between $G$-$Q$ and $\pi$-$N$ systems reminded me
of the old Fermi--Yang suggestion~\cite{FY49} of 1949,
 ``Are Mesons Elementary Particles?''
They proposed $\pi \sim N\bar N$ boundstate, 
the problem then is why $m_\pi^2 \ll (2m_N)^2$.
But this was taken care of by the Goldstone theorem.
So, when I realized Eq.~(22) is true, for a brief period,
I wondered whether the Gap Equation of Fig.~\ref{GapYuk}
could be applied to the $\pi$--$N$ system.
But then I realized that the $\pi$--$N$ system took the
path of QCD, that of stringy resonances or hadrons.
Soon afterwards, Fermi himself was diverted by the 
appearance of resonances, while for
the analogy to Fig.~\ref{GapYuk}, one could not integrate
$q^2$ up to $(2m_N)^2$, since around 500 MeV onwards,
mesons appeared.

Given $G \sim Q\bar Q$ and $\lambda_Q = \sqrt2 m_Q/v$,
and our numerical solution to Fig.~\ref{GapYuk},
i.e. dynamical mass generation for the structureless
or elementary $Q$, I wonder whether we could have
a second chance of realizing a non-QCD strong Yukawa 
version of the Fermi--Yang suggestion.
The Goldstone $G$ would be a deeply bound~\cite{Hou:2012az} 
$Q\bar Q$ state, while this illustrates that the strong Yukawa
Gap Equation approach cannot be the traditional TC, 
Walking or not.
Although we cannot offer deeper insight on this
underlying non-QCD theory, we suggest that
the true UV theory, likely containing the Theory
of Yukawa couplings, would contain Scale Invariance
violation.

\section{\underline{Conclusion}}

We have proposed a way of dynamical electroweak symmetry breaking
by heavy 4th generation ``bootstrap'':
the large Yukawa coupling $\lambda_Q$ of massless $G$ (or $V_L$) 
exchange between $Q$ and $\bar Q$ generates $m_Q$ dynamically,
which in turns assures the massless Goldstone nature of $G$ as input.
Although the term is from Nambu, we of course caution that
the only successful ``bootstrap'' is not lifting oneself
by the boots' straps, but breaking the boots and lifting up the
upper parts ...

Let us summarize the salient features.
\begin{itemize}
\item
An empirical-based dynamical Gap Equation by Goldstone, or $V_L$, exchange 
with aforementioned strong Yukawa coupling is constructed, 
and solved numerically. %\\ 

This can, in principle, replace usual Higgs field Condensation.
\item
The needed Yukawa coupling is of order $4\pi$!!
This is reminiscent of the $\pi$-$N$ system,
and implies 4G masses in 2-3 TeV Range. %\\

$Q\bar Q$ boundstates decaying via multi-$V_L$ could aid discovery.
\item
The 126 GeV Higgs-like boson poses difficulties for this picture.
For our perspective to remain viable, this would have to be a dilaton ...

Our Gap Equation is scale-invariant,
hence allows for a dilaton to appear.
\item
VBF production would arbitrate on the dilaton possibility,
but this would have to wait until 2015 or beyond.
\end{itemize}

Because of the importance of knowing the true source of EWSB,
let us wait what Nature reveals to us through the LHC.

\section{Addendum}

So, what about ``CP Violation for the Universe''?
We have focused only on EWSB as possibly induced by strong Yukawa coupling. 
But I have noticed~\cite{Hou:2008xd} a few years ago that,
if one extends SM to SM4, i.e. by adding 4G quarks,
then the Jarlskog invariant, the source of CPV in SM,
would jump by a factor of 1000 trillion or so,
and seem sufficient to bridge the needed gap of 10 billion.
This adds to the importance of keeping the
watch on whether Nature could still provide us with 4G.

\vskip0.3cm
\noindent\textbf{Acknowledgement}\
We thank Maria Garzelli for a comment that changed the
layout plan of this writeup, which is different from
how it was presented in Venice.
We thank Jean-Marc G\'erard for a pleasant visit to
Louvain, and Mathieu Buchkremer for the arrangements.

\end{document}